\pdfoutput=1
\documentclass[prd,twocolumn,eqsecnum,nofootinbib,floatfix]{revtex4}
\usepackage[titletoc,title]{appendix}
\usepackage{amsmath}
\usepackage{bm}
\usepackage{graphicx} 
\usepackage{float}
\newcommand{\E}{{\cal E}} 
\newcommand{\B}{{\cal B}}
\newcommand{\R}{{\cal R}}

\allowdisplaybreaks
\begin{document}
\title{Deflection of light by the field of a binary system} 
\author{Nahid Ahmadi} 
\affiliation{Department of Physics, University of Tehran, Kargar Avenue North, Tehran 14395-547, Iran} 
\author{Zainab Sedaghatmanesh } 
\affiliation{Department of Physics, University of Tehran, Kargar Avenue North, Tehran 14395-547, Iran} 
\affiliation{Department of Physics, Kharazmi University, Tehran, Iran}
%\date{January 2015} 
\begin{abstract} 
We examine the motion of a photon in the gravitational field of a binary system. The equations of motion are geodesic equations in a Schwarzchild background with a tidal force. We specialize the equations to that of an edge-on binary and use the method of osculating elements to integrate them. This work helps to identify a binary system through the gravitational light deflection of one member in the gravitational field of the other member. It is found that the effects of the companion body on a photon passing the edge of the star can be potentially detected by astrometric satellites with $\mu$as precision, if the ratio of the Schwarzchid radius to the star radius, $\frac{Gm}{c^2 R}\geq 10^{-5}$. Two different cumulative effects on the photon path are identified. 
\end{abstract} 
%\pacs{04.20.-q, 04.25.Nx; 04.70.Bw; 97.60.Lf}
\maketitle
%%%%%%%%%%%%%%%%%%%%%%%%%%%%%%%%%%%%%%%%%%%%%%%%%%%%%%%%%%%%%%%%%%%%%%%%%%%%%%%%%
\section{Introduction } 

How is the orbit of a photon that is affected by the field of a binary system? This is a three- body problem with gravitational interactions. Adding one body to the famous two- body problem of a photon in a gravitational field brings a remarkable complexity to the problem. It is however believed that in a complete treatment of such motions and with todays highly accurate astronomical observations in hand, these effects must be taken into account. This subject has been a field of interest since the astrometric space mission like GAIA planned to position the stars and celestial objects with one microarcsecond level of accuracy \cite{gaia}. At this level, many subtle effects are potentially observable. One of them would probably be the deflection of light by the field of a binary system. When a photon moves in the (weak gravity) field of an isolated spherical body, the leading order corrections to the photon (Newtonian path) are scaled by $\frac{Gm}{c^2 R}$, where $R$ is the closest distance to the gravitating body the photon can reach and $m$ is the mass of such body. Considering the scales characterizing our problem, it is tempting to ask whether there are any relativistic effects produced by the second body in the binary that we should be concerned of, or the traces left by the companying body still remains a challenge for future missions?   

In this work, we assume that photons move on the null paths and are influenced by a binary system, with two equal masses, in a circular motion. We also assume that the radius of the binary orbit is large, compared to the distance between photon and the member of the binary which we call it the {\it main body}. In adequately weak gravity regimes, effects like the deviation of the photon trajectory from straight line, may be well described by the post-Newtonian approximation. The latter assumption permits an approximate solution to our problem by reducing the three-body problem to the motion of a photon in an effective field of the orbiting bodies. 

To better discuss the dynamics of this system, we introduce two lengthscales. One scale is given by $M:=\frac{Gm}{c^2}$, and the other scale is the radius of the circular orbit, denoted by $r_o$. Let $\rho$ be the separation between the photon and the black hole. We assume that i) ${M}/{\rho}$ is small and ii) the second body in the binary is remote. In a reference frame that moves with the main body, the effects of the second body on the geometry of the spacetime can be described by tidal potentials. These are given by a linear perturbation about the Schwarzchild solution; so that the perturbed metric satisfies the vacuum field equations, perturbatively. 

It is straightforward to apply tidal effects to a system comprising a photon and a gravitating body, but solving the equations corresponding to the path of the photon in this perturbed spacetime would be tricky. A suitable framework for solving the path based on {\it osculating orbits} has already been suggested in \cite{Nahid}. In this framework, the motion of the photon is at all times described by a family of  null geodesics, called osculating trajectories, characterized by a set of constants of motion. A nice reformulation of the various aspects of the original problem can be achieved by the evolution of these constants, if we know the interacting forces. The evolution of this set of constants is constrained by a set of first order differential equations, known as the osculating conditions. In this paper we calculate the tidal forces acting on the photon employing a metric calculated in a paper by Taylor and Poisson \cite {Taylor Poisson}. In their work, the main gravitating body was a black hole, they used this metric to find the {\it dynamics of a black hole}, i.e, those produced by an influx of gravitational energy across the horizon (like tidal heating or increase in mass, angular momentum and surface area \cite{Poisson}) of the black hole. Our interest in this paper is mostly in the consequences of the tidal interactions on a {\it photon orbital dynamics}. The metric given by Taylor and Poisson is not necessarily restricted to the weak field approximation and can be equally used in small hole approximation, in which the gravitating body is assumed to be much smaller than the second body. In our problem, we need to know the tidal fields and their effects on the structure of spacetime around the gravitating body, all within the weak field approximation to general relativity. 

Assuming that the photon path would not be too different from the Schwarzchild light paths, we will include the force components in the osculating equations given in \cite{Nahid} to obtain the observable variations at any point of the path. Our results can be evaluated in any coordinate systems; this guarantees that those are observable quantities. For example, we find that the total deflection angle depends on the second body position and varies from a minimum equal to $4\frac{M}{R_*}\left[1-\pi\left(\frac{R_*}{r_o}\right)^3\right]$ to a maximum  $4\frac{M}{R_*}\left[1+2\pi\left(\frac{R_*}{r_o}\right)^3\right]$, where $R_*$ is the closest distance to the main body that the photon could reach, if it were an isolated body. 

The structure of this paper is as follows: In Section II, we start with the metric describing the tidal environment around a gravitating body. We find the tidally perturbed metric in Schwarzchid coordinates by a coordinate transformation (from harmonic coordinates). We would like to focus on the leading effects and neglect the smaller ones; so we find the simultaneously weak field and slow motion limit of the metric. These perturbations to the metric are considered as static fields on a flat background. We then calculate different components of the perturbing force and the equations governing the photon path in section III. In section IV, the evolution of constants of motion is discussed. The discussion will be concluded with the summary and conclusions of the results in section V. 
 %%%%%%%%%%%%%%%%%%%%%%%%%%%%%%%%%%%%%%%%%%%%%%%%%%%%%%%%%%
 \section{Tidal environment around a gravitating body}
  The general description of the spacetime around a spherical gravitating body in an arbitrary tidal environment is given by a metric, which has the following form in Schwarzchild coordinates, $\left(t,\rho, \theta^A=\left(\theta,\phi\right)\right)$, 
  \begin{widetext} 
\begin{subequations}
\label{background_metric_cart} 
\begin{align} 
g_{tt}&= -\left(1-2\frac{M}{\rho}\right)-\rho^2\left(1-2\frac{M}{\rho}\right)^2\E^q+O\left(\frac{\rho^3}{\R^2  \cal{L}}\right),  
\\
g_{\rho t} &= 0 ,\\
g_{At}&= \frac{2}{3}\rho^3\B^{q}_{a}\Omega^{a}_{A}+O\left(\frac{v}{c}\frac{\rho^3}{\R^2  \cal{L}}\right),\\
g_{\rho A}&= -\left[M\rho^3+\frac{1}{3}M^3\left(1-\frac{M}{\rho}\right)^{-2}\right]\E^{q}_a \Omega^{a}_{A}+O\left(\frac{\rho^3}{\R^2  \cal{L}}\right),\\
g_{\rho\rho}&=\left(1-2\frac{M}{\rho}\right)^{-1}-\rho^2\E^q+O\left(\frac{v}{c}\frac{\rho^3}{\R^2  \cal{L}}\right),\\
g_{AB}&=\rho^2\left[\Omega_{AB}-\rho^2\left(1-\frac{M}{\rho}\right)^{-1}\left(1-2\frac{M}{\rho}\right)^2\Omega_{AB}\E^q\right.\nonumber\\& \quad \mbox{} 
\left.+\left(\frac{1}{3}\frac{M^3}{\rho}\left(1-\frac{M}{\rho}\right)^{-1}-M\rho\left(1-\frac{M}{\rho}\right)\right)\left(2\E_{ab}\Omega^{a}_{A}\Omega^{b}_{B}+\E^q\Omega_{AB}\right)\right]+O\left(\frac{\rho^3}{\R^2  \cal{L}}\right)g_{ab}.
\end{align}
\end{subequations} 
\end{widetext}
 Here $M$ is gravitational radius of our body, $\Omega^{a}_{A}=\frac{\partial\Omega^a}{\partial\theta^A}$ which satisfies $\Omega^{AB}\Omega^{a}_{A}\Omega^{b}_{B}=\gamma^{ab}$ and $\Omega^{A}_{a}\Omega^a=0$. Furthermore, $\Omega^{A}_{a}=\Omega^{AB}\Omega_{a B}$, $\Omega^a:= \frac{x^a}{r}=\left(\sin\theta\cos\phi, \sin\theta\sin\phi,\cos\theta\right)$ is a unit radial vector, $\Omega_{AB}$ is the metric on the unit two-sphere, $x^a$'s are coordinates of a Cartesian system, and $\gamma_{ab}=\delta_{ab}-\Omega_a\Omega_b$ is a projection tensor on the direction orthogonal to $\Omega^a$. The domain of validity of the metric is the spherical body neighborhood. To find this metric, we used the results of \cite{Taylor Poisson}, and performed a coordinate transformation from the harmonics coordinates $\left(x^{0},x^a\right)$ to $\left(t,\rho, \theta,\phi\right)$, given by $x^0=ct, x^a=r\Omega^a\left(\theta^A\right)$, $r=\rho-M$. In \cite{Taylor Poisson}, Taylor and Poisson, examined the motion and the tidal dynamics of a black hole placed within a post-Newtonian external spacetime. Appendix A presents some crucial results of that paper. Our work is a continuation of their effort, and this coordinate transformation brings the metric of the distorted black hole (\ref{background_metric_harmonic}) to the new form given in (\ref{background_metric_cart}). This metric applies to the spacetime outside any spherical body immersed in a tidal environment. It is linearized about the Schwarzchild solution and satisfies (approximate) vacuum field equations. The tidal environment is presented as a functional of arbitrary external potentials $\left\{\E^{q}, \E^{q}_{a},\E^{q}_{ab},\B^{q}_{a} \right\}$ built from the lowest order tidal moments, $\E_{ab}$ and $\B_{ab}$. Within this approximation, the metric is accurate to all orders in $\frac{M}{\rho}$ and the neglected terms involve third order in $\frac{\rho}{r_o}$ that come from higher order tidal moments or the time derivative of quadrapole moments. Although the potentials that characterize the tidal perturbations are completely general, in this work we require that the gravitating body be a member of a binary system.
  
In studying the path of a photon through a curved geometry, it is a well known practice to consider the metric perturbations as fields defined on a flat background spacetime. In this way the tidal environment is described by the weak field limit of the metric. Following the same practice, we take $\frac{M}{\rho}$ as well as the potential terms produced by the second body, i.e the linear perturbation terms to flat Minkowski space, to be small. For the problem under study, the radius of the binary is large and the terms beyond the quadrapole tidal moments were already assumed to be small\footnote{The light defection in the post-linear gravitational field of two bounded masses for the case
when the impact parameter is much larger (5 times or more) than the distance between binary members, is calculated in \cite{Brugman}. In the following, we show that the linear terms are so small that we can confidently neglect the post-linear terms.}.
For this procedure to make sense, we should invoke the slow motion limit, $\frac{v}{c}=\sqrt{\frac{M}{r_o}}\ll 1$ as well. We introduce two dimensionless parameters which are almost equally small and denote them by $\epsilon$
\begin{equation} 
\epsilon:=\frac{M}{\rho}\approx\left(\frac{\rho}{r_o}\right)^4.
\label{epsilon} 
\end{equation} 
In this limit, the nonzero components of the metric are thus written as
\begin{subequations}
\label{background_weak field} 
\begin{align} 
g_{tt}&= -\left(1-2\frac{M}{\rho}+\rho^2\E^q\right)+O\left(\epsilon^2\right), \\
g_{\rho \rho} &= 1+2\frac{M}{\rho}-\rho^2\E^q +O\left(\epsilon^2\right),\\
g_{AB}&=\rho^2\Omega_{AB}\left[1-\rho^2\E^q+O\left(\epsilon^2\right)\right]
,\end{align}
\end{subequations} 
where $\E^q=\frac{1}{c^2}\E_{ab}\Omega^a\Omega^b$ and $\E_{ab}$ is a symmetric-trace free tensor, which its components for circular orbit motion  are calculated by matching the local metric with the global metric that include the gravitating body and the external spacetime \cite{Taylor Poisson}. These components are given in Appendix. By substituting (\ref{tidal moments}) in (\ref{tidal potentials}), the scalar potential, $\E^q$, is obtained as
\begin{equation} \label{scalar potential}
\E^q=A\left[1-3\cos^2\theta+3\sin^2\theta \cos 2\left(\phi-\psi\right)+O\left(\epsilon\right)\right],
\end{equation} 
where $A:=-\frac{M}{2r_{o}^3}$ is a dimensionful parameter and $\psi=\omega t$, where $\omega$ denotes the angular frequency of the tidal moments with respect to the frame moving with the main body. Before exploring the behavior of the photon in this spacetime, there are some points that we would like to add:
\begin{enumerate} 
\item Although the gravitomagnetic tidal (vector) potentials exist in a relativistic description of the geometry around a gravitating body, they do not appear in weak field limit and slow motion context ($v/c\ll 1$, where $v$ is the tidal environment velocity scale). This is also true for gravito-electric vector and tensor potentials. 
\item The spherical symmetry of the Schwarzchild geometry is perturbationally broken; so the Birkhoff theorem will let us to have a non-static vacuum spacetime.
\item The tidal geometry is described by a scalar potential and each scalar potential transforms as such under parity. This ensures that the deformed metric be conserved under a parity transformation. 
\item If we expand the tidal potential given in (\ref{scalar potential}), in terms of spherical harmonics
we will have \begin{equation}
\E^q=\sqrt{\frac{\pi}{5}}A \left[-Y^{0}_{2}+\sqrt{24}\left(Y^{2}_{2}+Y^{-2}_{2}\right)+O\left(\epsilon\right)\right].
\label{potential expansion} 
\end{equation} 
It is observed that the coefficients of $Y^{\pm1}_{2}$ are zero, and the metric components are invariant under $\theta\rightarrow\pi-\theta$. This fact demonstrates an additional symmetry with respect to the plane $\theta=\frac{\pi}{2}$.  The presence of the second body has greatly reduced the isometry (Lie) group of Schwarzchild spacetime to (discrete) subgroups, namely parity and mirror symmetry.
\end{enumerate} 
%%%%%%%%%%%%%%%%%%%%%%%%%%%%%%%%%%%%%%%%%%%%%%%%%%%%%%%%%%%%%%%%%%%%%%%%%%%%%%%%%%%%%%%%%%%%%
\section{null trajectories in the perturbed spacetime}
The first step before writing the null geodesic equations is to calculate the Christoffel symbols for this metric. The perturbed geodesic equations will then be written as
\begin{equation}\label{geodesics}
\frac{d^{2}x^{\mu}}{d\lambda^2}+\bar{\Gamma}^{\mu}_{\nu\sigma}\frac{dx^{\nu}}{d\lambda}\frac{dx^{\sigma}}{d\lambda}=F^\mu,
\end{equation} 
where
\begin{equation}\label{Gamma}
F^\mu=-\delta\Gamma^{\mu}_{\nu\sigma}\frac{dx^{\nu}}{d\lambda}\frac{dx^{\sigma}}{d\lambda}, \quad \mbox{}  \Gamma^{\mu}_{\nu\sigma}={\bar{\Gamma}}^{\mu}_{\nu\sigma}+\delta\Gamma^{\mu}_{\nu\sigma}.
\end{equation}
Here $\bar{\Gamma}$s and $\delta\Gamma$s are components of the Schwarzchild Christoffel symbols and their perturbations, respectively. Appendix B gives the full list of these symbols in the weak field limit. The right hand side of (\ref{geodesics}) represents the perturbing force components.

 It does not seem to be much hope for solving the set of equations (\ref{geodesics}) in a general case. Fortunately, there are some special cases in which the remaining symmetries simplify the task. It is easy to see that the plane $\theta=\frac{\pi}{2}$ satisfies the equation for $\theta\left(\lambda\right).$\footnote{If we calculate (\ref{geodesics}) for $x^\mu=\theta$, different terms include either $\Gamma^{\theta}_{\theta\mu}\dot{\theta}\dot{\mu}$ or $\Gamma^{\theta}_{\mu\nu}$ which can be easily verified from (\ref{symbols}) to be zero at $\theta=\pi/2$.} Therefore, if we specialize to the case in which photons move in equatorial plane of the binary system, i.e, if the inclination parameter, $i$, is exactly $\pi/2$ and the binary orbit is edge-on, the direction of angular momentum will be conserved and the photon will not leave the plane, because $F^\theta=0$. Other components of the tidal force applied to the photon will then be
\begin{subequations}\label {forces}\begin{align}
F^t&=\frac{1}{2}\rho^2\left(\rho^2\dot{\phi}^2+\dot{\rho}^2-\dot{t}^2\right)\partial_t\E^q 
\nonumber\\& \quad \mbox{} 
-2\rho\dot{t}\dot{\rho}\E^q-\rho^2\dot{t}\dot{\phi}\partial_\phi\E^q+O\left(\epsilon^2\right)\\
F^\rho &=\rho\left(\dot{\rho}^2-\dot{t}^2-\rho^2\dot{\phi}^2\right)\E^q
\nonumber\\& \quad \mbox{} 
+\rho^2\left(\dot{t}\partial_t\E^q+\dot{\rho}\partial_\phi\E^q\right)+O\left(\epsilon^2\right)\\
F^\phi &=\frac{1}{2}\left(\rho^2\dot{\phi}^2-\dot{\rho}^2-\dot{t}^2\right)\partial_\phi\E^q
\nonumber\\& \quad \mbox{} 
+\rho^2\partial_t\E^q+2\rho\dot{\rho}\dot{\phi}\E^q+O\left(\epsilon^2\right).
\end{align}\end{subequations}
Even with above simplifications, integrating the equations of motion and determining the photon path will not be simple. Although the perturbing forces are complicated, we believe that these forces keep the form of the photon path. Therefore, the leading correction to the Schwarzchild null curves can be determined by the method of Variation of Parameters (VoP). As described in \cite{Nahid}, in this method one employs a known solution of a fiducial system of differential equations as an ansatz for solving the system under study. The known solution, $x^{\mu}_{G}$, plays the role of instantaneously tangential (or osculating) trajectories to the true path, $x^\mu\left(\lambda\right)$. Each point on the true path with parameter $\lambda$ is coincides geometrically and dynamically with an osculating trajectory with one specified set of constants, $I^\alpha$; the number of these constants is equal to the number of phase space independent variables. The constants are furnished with $\lambda$ dependence and
%the true path is modeled {\it exactly} by a set of evolving constants. Photon motion may at all times be described by a set of $I^\alpha$, that 
evolve along a flow line on a cross section of phase space that is consistent with null condition. Each and every flow line is constrained by the osculating conditions
\begin{equation}\label {osculating conditions}
\frac{\partial \dot{x}^{\mu}_{G}}{\partial I^\alpha}\dot{I}^\alpha=F^\mu, \ \ \ \ \frac{\partial x^{\mu}_{G}}{\partial I^\alpha}\dot{I}^\alpha=0, \end{equation}
but different flow lines may correspond to very different interactions. This reformulation of the problem provides an insight into the effects of perturbing forces on various aspects of the motion. In the framework of VoP, the perturbing forces are not assumed to be small and the only restriction is that the forces must preserve the shape of the path. 

 Following \cite{Nahid}, these constants are called osculating elements, and the Schwarzchild null geodesics family, with parameter $\lambda$, is chosen as $x^{\mu}_{G}$. We recall that in weak field approximation, the relations describing {\it forced} Schwarzchild null geodesics are
\begin{subequations}\label {forced geodesics}\begin{align}
\rho^{-1}\left(\phi\right)&= \frac{\sin \left(\phi-\Delta\left(\phi\right)\right)}{R\left(\phi\right)}\nonumber\\& \quad \mbox{} +\frac{3M}{2R^2\left(\phi\right)}\left[1+\frac{1}{3}\cos2\left(\phi-\Delta\left(\phi\right)\right)\right],\label{geodesic rho}\\
\phi \left(\lambda\right)&=\Delta\left(\lambda\right)+\int^{\lambda}_{0} L\left(\tilde{\lambda}\right)\rho^{-2}\left(\tilde{\lambda}\right)d\tilde{\lambda},\\
t\left(\lambda\right)&=T\left(\lambda\right)+\int^{\lambda}_{0}\frac{E\left(\tilde{\lambda}\right)}{1-2M\rho^{-1}\left(\tilde{\lambda}\right)}d\tilde{\lambda},\\
\dot{\rho}^2&=E^2-\frac{L^2}{\rho^2}\left(1-\frac{2M}{\rho}\right)\label{rho geodesic}.
\end{align}\end{subequations}
Here dot denotes derivative with respect to $\lambda$ and the osculating elements on these trajectories are $\left\{E,L,R,\Delta,T\right\}$. Some osculating elements, like $R$, $T$ and $\Delta$, that characterize the initial position on the fiducial trajectory, are usually referred to as the positional elements; these elements typically change whenever the geometry of spacetime is affected with a new interaction. Other elements, like $L$ and $E$, which are called principal elements, determine on which trajectory the photon is moving. We can ignore $T$ and evolve $t$ explicitly by using 
\begin{equation} 
\acute{t}=\frac{E}{L}\frac{\rho^{2}}{1-2\frac{M}{\rho}},
\end{equation}
which is equivalent to the evolving of $t-T$, directly. The flow lines that satisfy (\ref{osculating conditions}) are given by
\begin{subequations}\label{osculating evolution}\begin{align}
\acute{E}&=\frac{R^{2}}{L}\frac{1}{\sin^2\left(\phi-\Delta\right)}F^{t}+O\left(\epsilon^2\right),\\
\acute{\Delta}&=-\frac{R^{3}}{L^2}\frac{1}{\sin\left(\phi-\Delta\right)}F^{\rho}+O\left(\epsilon^2\right),\\
\acute{L}&=\frac{R^{4}}{L}\frac{1}{\sin^4\left(\phi-\Delta\right)}F^{\phi}+O\left(\epsilon^2\right),\\
\acute{R}&=\frac{R^{4}}{L^2}\frac{\cos\left(\phi-\Delta\right)}{\sin^2\left(\phi-\Delta\right)}F^{\rho}+O\left(\epsilon^2\right),\\
\acute{t}&=\frac{E}{R^{2}L}\sin\left(\phi-\Delta\right)\left[\sin\left(\phi-\Delta\right)+4M\right]+O\left(\epsilon^2\right)\label{t evolution}.
\end{align}\end{subequations}
In these equations, $\phi$ has been used instead of $\lambda$ (or $t$), as an independent variable and prime denotes derivative with respect to $\phi$. The equations governing the evolution of $E$ and $L$ could have been calculated by the covariant formulation of osculating conditions \cite{Gair et al}.

The variations in the osculating elements are assumed to be small, therefore one can achieve a good approximation of these elements by substituting the leading order expressions, 
\begin{subequations}\label{geodesics at leading order}\begin{align}
\rho&=\frac{R_*}{\sin\left(\phi-\Delta_*\right)}\left(1+O\left(\epsilon\right)\right),\\
\dot{t}&=E_*\left(1+O\left(\epsilon\right)\right),\\
\dot{\rho}^2&=E_{*}^2-L_{*}^{2}\left(\frac{\sin^2\left(\phi-\Delta_*\right)}{R_{*}^2}+O\left(\epsilon\right)\right),\label{rho leading order}\\
\dot{\phi}&=\frac{L_{*}}{R_{*}^2}\sin^2\left(\phi-\Delta_*\right)+O\left(\epsilon\right),
\end{align}\end{subequations}
into the right hand side of equations (\ref{osculating evolution}). Here the set of constants that the photon trajectory could have assumed, if the main body were in a complete isolation, are denoted by $\left\{E_*,L_*,R_*,\Delta_*\right\}$.
The different components of the perturbing force in the plane $\theta=\pi/2$ has the form of
\begin{widetext}\begin{subequations}\label{perturbing forces}\begin{align}
F^t&=2E_{*}L_{*}A\left[\frac{1}{k}\cot\left(\phi-\Delta_*\right)\left[1+3\cos2\left(\phi-\psi\right)\right]+3\sin2\left(\phi-\psi\right)\right]+O\left(\epsilon^2\right),\\
F^{\rho}&=-2E_{*}L_{*}A\left[k\cos\left(\phi-\Delta_*\right)\left[1+3\cos2\left(\phi-\psi\right)\right]+3\cos\left(\phi-\Delta_*\right)\cos2\left(\phi-\psi\right)\left[\frac{E_* R_{*}^2\omega}{\sin^2\left(\phi-\Delta_*\right)}-L_*\right]\right]+O\left(\epsilon^2\right),\\
F^{\phi}&=2E_{*}^2A \left[-k\left[1+3\cos2\left(\phi-\psi\right)\right]\sin2\left(\phi-\Delta_*\right)+3\sin2\left(\phi-\psi\right)\left[1+\frac{L_*}{E_*}\omega-k^2\sin^2\left(\phi-\Delta_*\right)\right]\right]+O\left(\epsilon^2\right)
.\end{align}\end{subequations}\end{widetext} 
  Here $k:={E_{*}R_*}/{L_*}$ is a dimensionless parameter, which its value can be found by substituting, $\rho=R_*$ in (\ref{rho geodesic}), $k=1+O\left(\epsilon\right)$. In the derivation of the force components, we have used the square root of (\ref{rho leading order}) with a minus sign, $\dot{\rho}=-E_*\cos\left(\phi-\Delta_*\right)+O\left(\epsilon\right)$. 
  
We recall that in preceding equations, $\psi$ is a function of $t$; so formally, its explicit dependence on $\phi$ must be considered. However, these equations will turn out to be simpler by recognizing that in the problem under study, osculating elements undergo two types of changes. The first is an oscillation with a period equal to the period of tidal moments (or multiple of it); the second is a steady drift in $\phi$. In other words, we have two timescales; $t$ and $\omega^{-1}$. On short timescale, $t\approx M$, the photon moves on a geodesy of the Schwarzchild spacetime with a {\it static} tidal field, characterized by $\left\{E,L,R,\Delta\right\}$. Over longer timescale, $\omega^{-1}\approx\sqrt{\frac{M}{r_{o}^3}}\approx\frac{M}{\epsilon^{\frac{15}{8}}}$, the displacement of the second body changes $\psi$ and causes the path to evolve. It is clearly desirable to compute an accurate trajectory for the whole domain of the validity of $\phi$ (or $t$). However, the linear perturbation method is limited to producing a model for the trajectory over rapid timescale, (or an snapshot of the trajectory,) in which the trajectory is described by the deviations from the pivot elements, $\left\{E_*,L_*,R_*,\Delta_*\right\}$, and the new configurations (of the second body) are neglected. Such approximations fall out of phase with the true path after the longer timescale $\approx \epsilon^{\frac{-15}{8}}M $. In fact a rigorous prescription for the path can be provided by a two-timescale analysis, but considering the order of accuracy by which the short term secular changes are captured, at the cost of discarding long term effects, it is sensible to assume that $\dot{\psi}\ll\dot{\phi}$ or
\begin{equation}
\frac{\omega\dot{t}}{\dot{\phi}}\approx\frac{R_{*}^{2}E_*\omega}{L_*}\frac{1}{\sin^2\left(\phi-\Delta_*\right)}\ll1.
\end{equation}

After invoking this approximation, the components of the perturbing force can be written as 
\begin{subequations}\begin{align}
F^t&=2E_*L_*A\cot\left(\phi-\Delta_*\right)\left[3\frac{\cos \left(\phi-2\psi+\Delta_*\right)}{\cos\left(\phi-\Delta_*\right)}+1\right],\\
F^{\rho}&=2E_*L_*A\sin\left(\phi-\Delta_*\right)\left[3\frac{\sin \left(\phi-2\psi+\Delta_*\right)}{\sin\left(\phi-\Delta_*\right)}-1\right],\\
F^{\phi}&=2E_{*}^2A\cos\left(\phi-\Delta_*\right)\left[3\frac{\sin \left(\phi-2\psi+\Delta_*\right)}{\sin\left(\phi-\Delta_*\right)}-1\right].
\end{align}\end{subequations} 
The evolution of the conserved quantities will then be reproduced by substituting the preceding equations into (\ref{osculating evolution}) as
\begin{subequations}\label{delta-E-L-R evolution} \begin{align}
\acute{\Delta}&=-2AR_{*}^{2}\left[3\frac{\sin \left(\phi-2\psi+\Delta_*\right)}{\sin\left(\phi-\Delta_*\right)}-1\right]\label{delta evolution}, \\
\frac{\acute{R}}{R_*}&=2AR_{*}^{2}\cot\left(\phi-\Delta_*\right)\left[3\frac{\sin \left(\phi-2\psi+\Delta_*\right)}{\sin\left(\phi-\Delta_*\right)}-1\right]\label{R evolution},\\
\frac{\acute{E}}{E_*}&=2AR_{*}^{2}\frac{\cos\left(\phi-\Delta_*\right)}{\sin^3\left(\phi-\Delta_*\right)}\left[3\frac{\cos \left(\phi-2\psi+\Delta_*\right)}{\cos\left(\phi-\Delta_*\right)}+1\right]\label{E evolution},\\ \frac{\acute{L}}{L_*}&=2AR_{*}^{2}\frac{\cos\left(\phi-\Delta_*\right)}{\sin^3\left(\phi-\Delta_*\right)}\left[3\frac{\sin \left(\phi-2\psi+\Delta_*\right)}{\sin\left(\phi-\Delta_*\right)}-1\right]\label{L evolution}. 
\end{align}\end{subequations}
It is easily seen that the corrections imposed to the Schwarzchild null paths by the second body are of order $AR_{*}^2$, that is $O\left(\epsilon^{\frac{7}{4}}\right)$. These corrections are so small that, even for photons with large impact parameters, we can neglect higher order terms in (\ref{background_weak field}) with confidence. These equations appear to be singular at $\phi=\Delta_*,\pi+\Delta_*$. The singularities correspond to the fact that Schwarzchild geodesics loose their geometric meaning for asymptotic values of $\phi$.

In the following section, we will give expressions for osculating elements; once these functions are known, the path is obtained from  equations (\ref{forced geodesics}).  Considering the long separation of the two timescales, we assumed that the geometry around the main body has time translation symmetry. Before concluding this section, we are interested in using this symmetry to calculate energy of the perturbed system, $\bar{E}$, as a constant on the path. The Killing vector whose conserved quantity is energy, is $K=\frac{\partial}{\partial t}$. The conserved energy is then given by 
\begin {equation}\label{total E}
\bar{E}=-p_\mu K^\mu =\dot{t}\left(1-\frac{2M}{\rho}+\rho^2\E^q+O\left(\epsilon^2\right)\right)
 \end{equation}
This is a perturbationally constant quantity even for the paths outside of the equatorial plane. Moreover, for the trajectories constrained to the equatorial plane, the {\it direction} of angular momentum is also conserved.
%%%%%%%%%%%%%%%%%%%%%%%%%%%%%%%%%%%%%%%%%%%%%%%%%%%%%%%%%%%%%%%%%%%%%%%%%%%%%%%%%%%%%%%%%
\section{Evolution of osculating constants}
\begin{figure}
\includegraphics[width=0.40\textwidth]{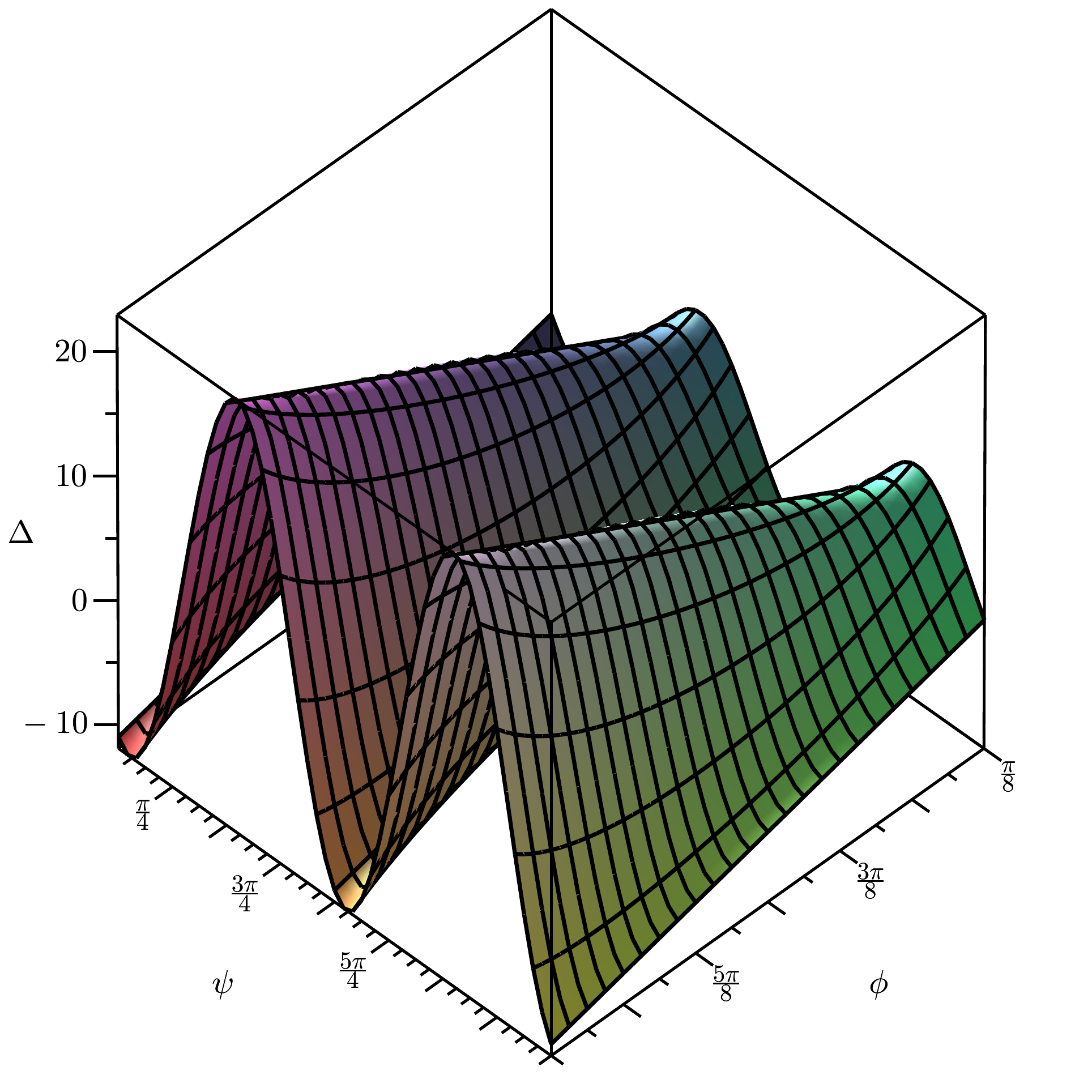}
\caption{Variations of $\Delta-\Delta_*$ vs. $\phi$ and $\psi$. Here $\Delta_*$ is assumed to be zero. On the vertical axis, we have chosen $AR_{*}^{2}=1$}
	\label{fig:plot3d5}
\end{figure}

The expression for the evolution of $\Delta$ in (\ref{delta evolution}) can now be used to obtain the explicit dependence on $\psi$ at each $\phi$. This gives
 \label{point deflection} \begin {align}
 \Delta\left(\phi\right)&=\Delta_*+2AR_{*}^{2}\left[\phi\left(1-3\cos2\left(\psi-\Delta_*\right)\right)\right.\nonumber\\
 & \quad \mbox{} +\left.3\sin2\left(\psi-\Delta_*\right)\ln\left(\sin\left|\Delta_*-\phi\right|\right)\right].
 \end{align}
 Fig.~\ref{fig:plot3d5} shows the variations of $\Delta$ at different points of the path for different $\psi$ directions at the initial condition $\Delta_*=0$. The contribution of the second body in the total deflection angle of a light ray passing near a binary system is
\begin{equation}\label{second body Delta deflection} 
\int^{\Delta_*+\pi}_{\Delta_*}\acute{\Delta} d\phi=2\pi AR_{*}^{2} \left[1-3\cos2\left(\psi-\Delta_*\right)\right]
.\end{equation}
This implies that for edge-on binaries, the second body gives maximum increase in the deflection angle when $\psi-\Delta_{*}=\left\{\frac{\pi}{2},\frac{3\pi}{2}\right\}$ and maximum decrease is obtained when $\psi-\Delta_{*}=\left\{0,\pi\right\}$. In \cite{Zschoke}, it is shown that among orbits with arbitrary inclination, the maximal possible light deflection is attained for edge-on binaries, ($i=\pi/2$). Our calculations complement the criteria for orbital elements by using those the maximal changes in positional elements are determined. The various gravitational effects in the light propagation are additive in weak field limit; so for the total deflection angle, $\delta$, we will find that
 \begin{equation}\label{total deflection} 
4\frac{M}{R_*}-4\pi AR_{*}^{2} \leq\delta\leq 4\frac{M}{R_*}+8\pi AR_{*}^{2}.
\end{equation}
 By integrating (\ref{R evolution}), one can obtain
 \label{R deflection} \begin {align}\frac{R}{R_*}&=1+2AR_{*}^{2}\left[6\sin 2\left(\Delta_*-\psi\right)\left(1-\cot\left(\phi-\Delta_*\right)\right)\right.\nonumber\\& \quad \mbox{}
  -\left.\left(1-3\cos 2 \left(\Delta_*-\psi\right)\right)\ln\sin^2\left(\phi-\Delta_*\right)\right]. \end{align}
 \begin{figure}
	\includegraphics[width=0.40\textwidth]{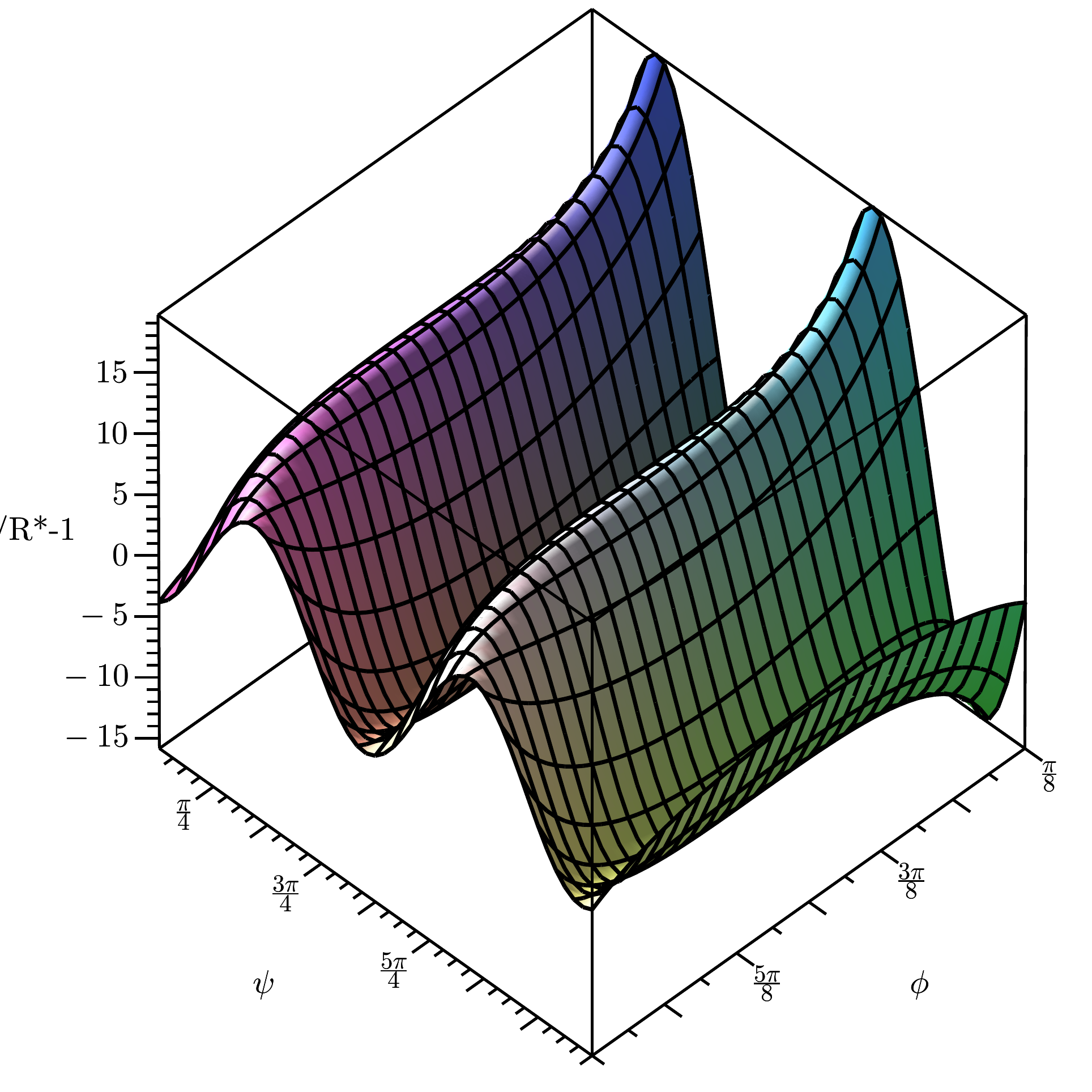}
	\caption{Variations of $R/R_*-1$ vs. $\phi$ and $\psi$. On the vertical axis, we have chosen $AR_{*}^{2}=1$}
	\label{fig:plot3d3}
\end{figure}
This amounts to a total change in $R$ equal to 
\begin{equation}\label{second body R deflection} 
12\pi AR_{*}^{3}\sin2\left(\Delta_*-\psi\right).\end{equation}
The photon gets a maximum value of total $R$- deflection when $\psi-\Delta_{*}=\left\{\frac{\pi}{4},\frac{3\pi}{4},\frac{5\pi}{4},\frac{7\pi}{4}\right\}$; no change in total $R$ is expected when $\psi-\Delta_{*}=\left\{\frac{\pi}{2},\pi,\frac{3\pi}{2}\right\}$.

\begin{figure}
\includegraphics[width=0.40\textwidth]{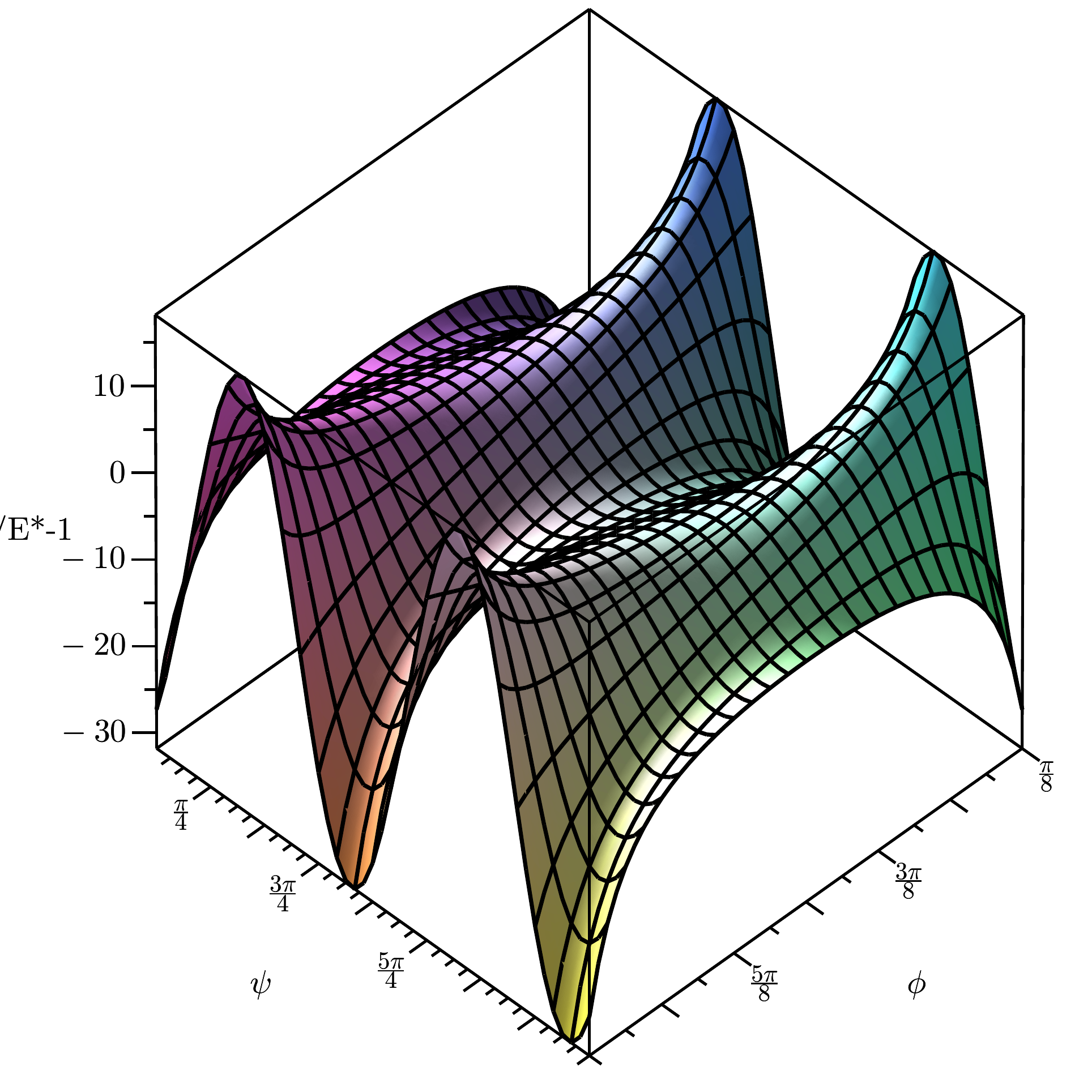}
	\caption{Variations of $E/E_*-1$ vs. $\phi$ and $\psi$. On the vertical axis, we have chosen $AR_{*}^{2}=1$}
	\label{fig:plot3d1}
\end{figure}

\begin{figure}
	\includegraphics[width=0.40\textwidth]{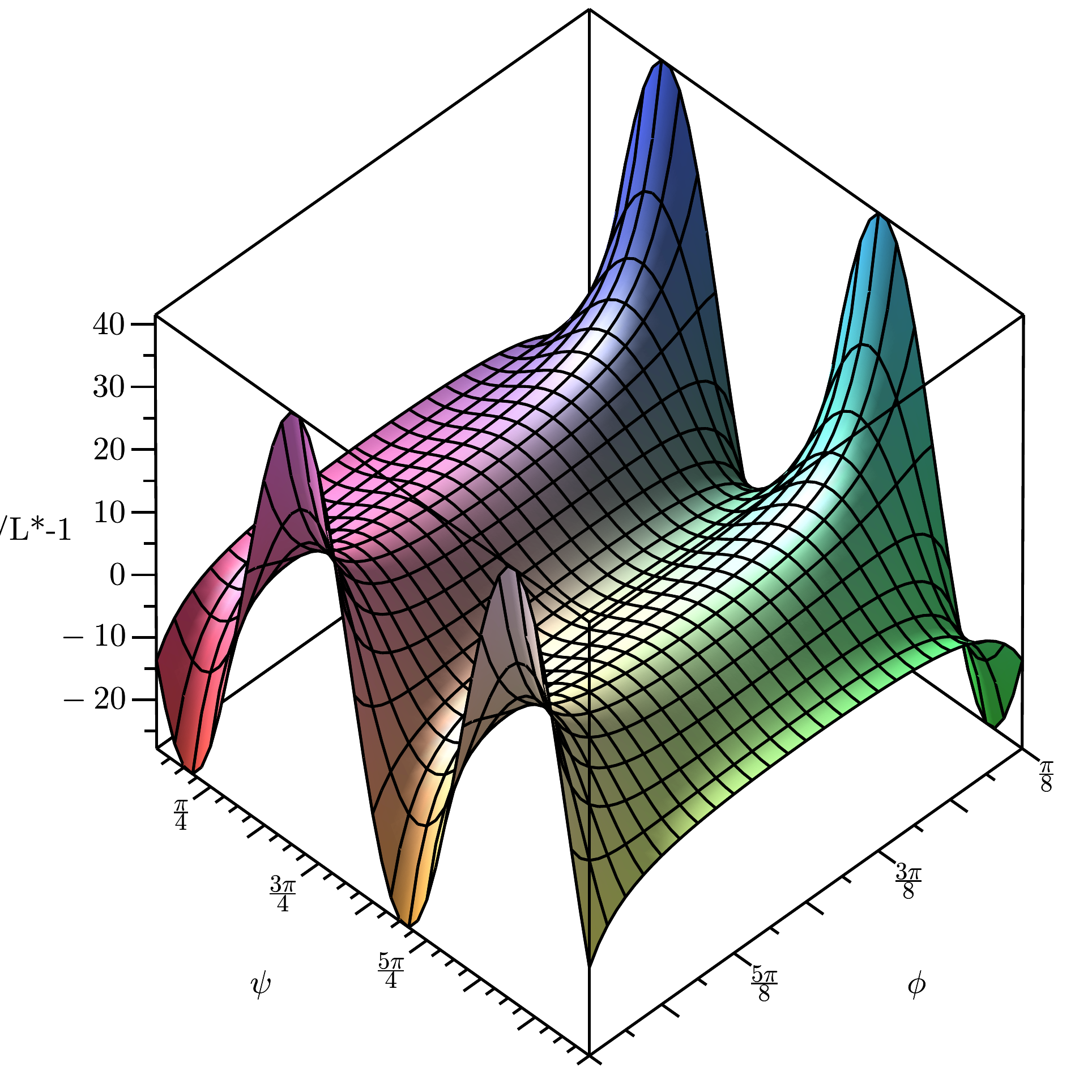}
\caption{Variations of $L/L_*-1$ vs. $\phi$ and $\psi$. On the vertical axis, we have chosen $AR_{*}^{2}=1$}
\label{fig:plot3d7}
\end{figure}

Obviously, one can find a different parametrization of the path by first integrating (\ref{t evolution}) and keeping the leading order terms to produce, 
\begin{equation}\label{t deflection} 
t\left(\phi\right)=\frac{E_*}{L_*R_{*}^{2}}\left[\frac{1}{2}\phi-\frac{1}{4}\sin2\left(\phi-\Delta_*\right)+4M\cos\left(\phi-\Delta_*\right)\right],
\end{equation}
and then omitting $\phi$ in favor of $t$, numerically. The expressions describing the evolution of other osculating elements, up to the leading order in $\epsilon$, are
\begin{subequations}\label{E-L-R} \begin{align}
\frac{E}{E_*}=1+AR_{*}^{2}&\left[6\sin 2\left(\Delta_*-\psi\right)\cot\left(\phi-\Delta_*\right)\right.\nonumber\\& \quad \mbox{}\left.
  -\frac{1+3\cos 2 \left(\Delta_*-\psi\right)}{\sin^2\left(\phi-\Delta_*\right)}\right]\label{E},\\ 
 \frac{L}{L_*}=1+AR_{*}^{2}&\left[-6\sin 2\left(\Delta_*-\psi\right)\cot^3\left(\phi-\Delta_*\right)\right.\nonumber\\& \quad \mbox{}\left.
  +\frac{1-3\cos 2 \left(\Delta_*-\psi\right)}{\sin^2\left(\phi-\Delta_*\right)}\right]\label{L}.  
\end{align}\end{subequations}
In Figs.~\ref{fig:plot3d5}-~\ref{fig:plot3d7}, we show the evolution of the path under the influence of the second body and initial condition $\Delta_*=0$. To avoid the singularities at asymptotes, $\phi=0,\pi$, the $\phi$ variations does not cover the whole domain, also we have chosen $AR_{*}^{2}=1$. The four figures show the two positional elements, $\left(\Delta,R\right)$, and the two principal elements, $\left(E, L\right)$, as functions of $\phi$ and $\psi$. The comparison of behavior of these elements at asymptotes shows that there is a secular change in $\left(\Delta,R\right)$; but there is no changes in the values of the principal elements in these limits. For particles moving on timelike bound orbits, it is believed that the changes in principal elements occur when perturbing forces include a dissipative part \cite{Pound Poisson}. Likewise, equations (\ref{L}) and (\ref{E}) imply that  $L\left(0\right)=L\left(\pi\right)$ and $E\left(0\right)=E\left(\pi\right)$ for massless particles; even though the total angular momentum is not conserved in our problem. Although this result is obtained for a tidal interaction and edge-on binaries, it seems to be also valid for general inclinations.
\section{summary and conclusions}
In this paper, we considered a photon in the gravitational field of a binary system and found the form of trajectories invoking the weak field and slow motion approximations. In practice, we found the magnitude of secular changes to the path of a light ray propagating in the field of a spherical gravitating body, that arises in the presence of a companion body. This work illustrates the application of the osculating elements formalism to the computation of photons paths evolutions in the the Schwarzchild spacetime, as well as the tidal geometry around a black hole in reducing a photon-binary problem to an effective two body problem.  
This work shows how a companion star can affect the curvature of a light ray brought about by the main lens. It is known that edge-on binaries deflect the light more than others \cite{Zschoke}; this criterion is further developed in this paper. Provided that the companion body is in a right distance and at a right orbital phase, in principle, it can be traced with a sensitive detector. The right distance, $r_o$, and phase, $\psi-\Delta_*$, are determined by the geometry of spacetime around the main lens. Equation (\ref{delta-E-L-R evolution}) indicates that edge-on binaries with $r_o=O\left(\epsilon^{-\frac{5}{4}}M\right)$, and $\psi-\Delta_{*}=\left\{\frac{\pi}{2},\frac{3\pi}{2}\right\}$ can produce deflections of order $O\left(\epsilon^{\frac{7}{4}}\right)$. Therefore, these corrections have potentially detectable effects on GAIA (or any other $\mu$as $\approx10^{-8}$ astrometric mission) observables, if $\epsilon\geq10^{-5}$. This condition holds for known typical binary systems. In this estimation, it is assumed that the photon passes by the edge of the star.

The importance of determining such corrections to Schwarzchild null geodesics can be understood when astrometric information is read off the light paths at a high precision level. The astrometric observations of a GAIA-like satelite has been modeled within the PPN formalism to Post-Newtonian gravity. The sensitivity of such astronomy to some PPN parameters \cite{Kopeikin} as well as the light bending due to quadrapole moment of an axisymmetric planet (like Jupiter) \cite{Crosta} has been established. In such feasibility studies, one must consider the light deflection in the vicinity of a star which is i) a member of a binary system and ii) crosses one of the astrometric fields during the GAIA's mission. A comparision between the details of the background field corresponding to the time when the star is at the center of the field with that when it is not, will help to extract the contribution of the binary system. The difference between the two patterns may then be fitted to the deflection model to determine whether the effect is detectable or not. In  spite of the common belief that 50 percent of stars are a member of a binary (or multiple) system \cite{Duquennoy} and the order of correction terms is in GAIA's accuracy level, finding a favourable field and binary system with known ephemerides is a big challenge that is out of the scope of this paper. On the other hand, the trace of wide binary lenses studied in this paper may be found in microlensing light curves. If repeating microlensing peaks \cite{Mao} appear in the microlensing light curves during the GAIA's mission period, it will be a good signal for a binary candidate with $ M\ll r_o$.
Analytical results derived in this paper can be used in models fitted to the microlensing light curves. Such models provide a convenient framework for mutual verification of the results in an astrometric field common in different sky missions. 

%%%%%%%%%%%%%%%%%%%%%%%%%%%%%%%%%%%%%%%%%%%%%%%%%%%%%%%%%%%%%%%%%%%%%%%%%%%%%%%%%%%%
\begin{appendices}\numberwithin{equation}{section}
\section{Tidal environment around the main body}
In this appendix we list some results presented in \cite{Taylor Poisson} which investigated a tidally deformed spacetime around a nonrotating black hole. The tidal environment around a slowly rotating black hole has been studied recently by Poisson \cite{Poisson- rotating}. To capture the effects associated with the dragging of inertial frames, this calculation required to go beyond first order perturbation theory. When the rotational deformations are applied to a circular binary, the tidal moments will incorporate terms that involve powers of $v/c$. Since these two metrics are the same at the orders that we require, the higher order terms do not affect on the results of our work. Some notations we use in our work, and are adapted from \cite{Taylor Poisson}, will be introduced here. We will express the results in the form required for our specific purpose.  

The metric components of the black hole immersed in an arbitrary tidal environment, in harmonic coordinates $\left(x^{0},x^a\right)$, are given by
 \begin{widetext} 
\begin{subequations}
\label{background_metric_harmonic}\begin{align}
g_{00}&=-\frac{1-\frac{M}{r}}{1+\frac{M}{r}}-r^2\left(1-\frac{M}{r}\right)^2\E^q+O\left(\frac{r^3}{\R^2  \cal{L}}\right),\\
g_{0a}&=\frac{2}{3}r^2\left(1-\frac{M}{r}\right)\left(1+\frac{M}{r}\right)^2\B^{q}_{a}+O\left(\frac{v}{c}\frac{r^3}{\R^2  \cal{L}}\right),\\
g_{ab}&=\frac{1+\frac{M}{r}}{1-\frac{M}{r}}\Omega_a\Omega_b+\left(1+\frac{M}{r}\right)^2\gamma_{ab}-r^2\left(1+\frac{M}{r}\right)^2\E^q \Omega_a\Omega_b
-Mr\left(1+\frac{M}{r}\right)^2\left(1+\frac{M^2}{3r^2}\right)^2\left(\Omega_a\E^{q}_{b}+\E^{q}_{a}\Omega_b\right)\nonumber\\
& \quad \mbox{} -r^2\left(1-\frac{M}{r}\right)^2\left(1+\frac{M}{r}\right)^3\gamma_{ab}\E^q
-Mr\left(1+\frac{M}{r}\right)^2\left(1-\frac{M^2}{3r^2}\right)\E^{q}_{ab}+O\left(\frac{r^3}{\R^2 \cal{L}}\right).
\end{align}
\end{subequations} 
\end{widetext} 
 Here $M$ is the gravitational radius of the black hole, when it is in complete isolation, $x^a$ are a system of Cartesian coordinates, $\Omega^a:= \frac{x^a}{r}=\left(\sin\theta\cos\phi, \sin\theta\sin\phi,\cos\theta\right)$ is a vector which satisfies $\delta_{ab}\Omega^a\Omega^b=1$ and $\gamma_{ab}=\delta_{ab}-\Omega_a\Omega_b$ is a projection tensor on the direction orthogonal to $\Omega^a$. The contribution of the tidal environment is expressed as an expansion in the strength of the tidal coupling. The expansion parameter is the inverse of the radius of curvature, $\R^{-1}$, of the external spacetime in which the black hole moves. In the neglected terms, $\cal{L}$ is the scale of spatial inhomogeneity in the external spacetime. At the lowest order in $\R^{-2}$, the metric can be described by a set of external potentials $\left\{\E^{q}, \E^{q}_{a},\E^{q}_{ab},\B^{q}_{a} \right\}$ built from quadrapole moments, $\E_{ab}$ and $\B_{ab}$ by 
\begin{subequations} 
\label{tidal potentials} 
\begin{align}
\E^q &= \frac{1}{c^2}\E_{ab}\Omega^a\Omega^b,\\
\E^{q}_{a}&= \frac{1}{c^2}\gamma_{a}^{\ c}\E_{cd}\Omega^d,\\
\E^{q}_{ab}&= \frac{1}{c^2}\left(2\gamma_{a}^{\ c}\gamma_{b}^{\ d}\E_{cd}+\gamma_{ab}\E^q\right),\\
\B^{q}_{a}&= \frac{1}{c^3}\epsilon_{apq}\Omega^{p}\B^{q}_{c}\Omega^{c}. 
\end{align}
\end{subequations} 
The tidal moments are formally defined in terms of the Weyl tensor of the external spacetime. In practice, however, for every application those are determined by matching the local metric to the global metric that includes the black hole and the external spacetime. For the specialized case that the black hole is part of a binary system in circular motion and the global spacetime is expressed in the post-Newtonian gravity regime, such a matching were carried out in \cite{Taylor Poisson} and \cite{Johnson-McDaniel et al}. For the problem under study in our paper where the two gravitational bodies in the binary system have equal masses and the relative orbital velocity, $v$, within the binary system is slow, $\frac{v}{c}=\sqrt{\frac{2M}{r_o}}=O\left(\epsilon^{\frac{5}{8}}\right)$, the nonvanishing components of the tidal moments are given by  
\begin{subequations}\label{tidal moments}\begin{align}
\E_{11}&=-\frac{M}{2r_{o}^3}\left[1+
3\cos2\psi+O\left(\frac{v}{c}\right)^2\right],\\
\E_{12}&=\E_{21}=-\frac{M}{2r_{o}^3}\left[1+
3\sin2\psi+O\left(\frac{v}{c}\right)^2\right],\\
\E_{22}&=-\frac{M}{2r_{o}^3}\left[1-
3\cos2\psi+O\left(\frac{v}{c}\right)^2\right],\\
\E_{33}&=\E_{11}+\E_{22}=-\frac{M}{r_{o}^3}+O\left(\frac{v}{c}\right)^2,\\
\B_{23}&=\B_{31}=-\frac{3M}{r_{o}^3}v\sin\psi+O\left(c^{-2}\right),\\
\B_{13}&=\B_{32}=-\frac{3M}{r_{o}^3}v\cos\psi+O\left(c^{-2}\right).
\end{align}\end{subequations}
Here $\psi$ is the phase related to the angular velocity, $\omega$, of the tidal moments with respect to the frame moving with the black hole, $\psi=\omega t=\sqrt{\frac{2M}{r_{o}^3}}t\left(1+O\left(\frac{v}{c}\right)^2\right) $.
 It can also be defined in terms of the orbital velocity of the second body, $\bar{\omega}$, and the precessional angular frequency of the black hole frame relative to the barycentric frame, $\Omega$, by $\psi=\bar{\omega}\bar{t}-\Omega t$, in which $\bar{t}$ is the global time coordinate.
%%%%%%%%%%%%%%%%%%%%%%%%%%%%%%%%%%%%%%%%%%%%%%%%%%%%%%%%%%%%%%%%%%%%%%%%%%%%%%%%%%%%
\section {Christoffel symbols}
The nonvanishing Christoffel symbols built from metric (\ref{background_weak field}) are
\begin{widetext}\begin{align}\label{symbols}\begin{tabular}{cccc}
\begin{tabular}{l}
 $\Gamma^{t}_{tt}= -\Gamma^{t}_{\rho\rho}=\frac{1}{2}\rho^2\frac{\partial}{\partial t}\E^q,$ \\ 
 $\Gamma^{t}_{t\rho}= \frac{M}{\rho^2}+\rho\E^q, $  \\
 $\Gamma^{t}_{tA}= \frac{1}{2}\rho^2\frac{\partial}{\partial \theta^A}\E^q,$\\
 $\Gamma^{t}_{AB}= -\frac{1}{2}\rho^4\Omega_{AB}\frac{\partial}{\partial t}\E^q, $ \\
  %\hline 
  \end{tabular}&
  \begin{tabular}{l} 
$\Gamma^{\rho}_{tt}= -\Gamma^{\rho}_{\rho \rho}= \frac{M}{\rho^2}+\rho\E^q,$\\ 
$\Gamma^{\rho}_{\rho t}= \frac{1}{2}\rho^2\frac{\partial}{\partial t}\E^q, $\\  
$\Gamma^{\rho}_{\rho A}=-\frac{1}{2}\rho^2\frac{\partial}{\partial \theta^A}\E^q, $ \\
$\Gamma^{\rho}_{AB}=\rho^3\Omega_{AB}\frac{\partial}{\partial t}\E^q,$ \\ 
%	\hline 
	\end{tabular}&
	  \begin{tabular}{l} 
$\Gamma^{\theta}_{tt}= -\Gamma^{\theta}_{\rho \rho}= \frac{1}{2}\rho\frac{\partial}{\partial \theta}\E^q, $\\
$\Gamma^{\theta}_{\theta t}=-\frac{1}{2}\rho^2\frac{\partial}{\partial t}\E^q,$ \\
$\Gamma^{\theta}_{\theta \rho}=\frac{1}{\rho}-\rho\E^q, $\\
$\Gamma^{\theta}_{\theta A}=-\frac{1}{2}\rho^2\frac{\partial}{\partial \theta^A}\E^q,$   \\
$\Gamma^{\theta}_{\phi \phi}=-\sin\theta\cos\theta+\frac{1}{2}\rho^2 \sin\theta\frac{\partial}{\partial \theta}\E^q, $ \\ 
	%\hline
	 \end{tabular}&
	 \begin{tabular}{l} 
$\Gamma^{\phi}_{tt}= -\Gamma^{\phi}_{\rho \rho}= \frac{1}{2\sin^2\theta}\frac{\partial}{\partial \phi}\E^q, $\\
$\Gamma^{\phi}_{\phi t}=-\frac{1}{2}\rho^2\frac{\partial}{\partial t}\E^q,$ \\
$\Gamma^{\phi}_{\phi \rho}=\frac{1}{\rho}-\rho\E^q,$\\
$\Gamma^{\phi}_{\theta \theta}=\frac{1}{2}\rho^2 \sin^2\theta\frac{\partial}{\partial \phi}\E^q, $ \\
$\Gamma^{\phi}_{\phi \phi}=-\frac{1}{2}\rho^2 \frac{\partial}{\partial \phi}\E^q.$\\
\end{tabular}\\
\end{tabular}
\end{align}
\end{widetext}
\end{appendices}
%%%%%%%%%%%%%%%%%%%%%%%%%%%%%%%%%%%%%%%%%%%%%%%%%%%%%%%%%%%%%%%%%%%%%%%%%%%%%%%%%%%%%%%%%%

\begin{thebibliography}{100}
\bibitem{gaia}
URL: http://sci.esa.int/gaia.
\bibitem{Nahid}
N. Ahmadi, Phys. Rev. D\textbf{80},124028 (2009).
\bibitem{Taylor Poisson}
S. Taylor and E. Poisson, Phys. Rev. D\textbf{78}, 084016 (2008). 
\bibitem{Poisson}
E. Poisson, Phys. Rev. D\textbf{70}, 084044, (2004); K. Alvi, Phys. Rev. D\textbf{64}, 104020 (2001).
\bibitem{Brugman}
M. H. Brugman, Phys. Rev. D\textbf{72}, 024012 (2005);
\bibitem{Gair et al}
Jonathan R. Gair, Eanna E. Flanagan, Steve Drasco, Tanja Hinderer and Stainslav Babak, Phys. Rev. D\textbf{83}, 044037 (2011).
\bibitem{Zschoke}
Sven Zschoke, The Astronomical Journal \textbf{144}, 77 (2012).
\bibitem{Pound Poisson}
Adam Pound and Eric Poisson, Phys. Rev. D\textbf{77}, 044013 (2008).
\bibitem{Poisson- rotating}
E. Poisson, Phys. Rev. D\textbf{91}, 044004 (2015).
\bibitem{Johnson-McDaniel et al}
N. K. Johnson-McDaniel, N. Yunes, W. Ticky and B. J. Owen, Phys. Rev. D\textbf{80}, 124039 (2009).
\bibitem{Kopeikin}
S. M. Kopeikin and V. V. Makarov, Phys. Rev. D\textbf{75}, 062002, (2007). 
\bibitem{Crosta}
M. T. Crosta and F. Mignard, Class. Quant. Grav.\textbf{23}, 4853, (2006). 
\bibitem{Duquennoy}
A. Duquennoy and M. Mayor, A\&A, \textbf{248}, 485 (1991); J. L. Halbwachs, M. Mayer, S. Udry and F. Arenou, A\&A \textbf{397}, 159 (2003).
\bibitem{Mao}
S. Mao, Research in Astronomy and Astrophysics, volume \textbf{12}, Issue \textbf{8}, 947 (2012); J. Skowron, L. Wyrzykowski, S. Mao and M. Jarosaynski, Mon. Not. R. Astron. Soc., 393, 999 (2009); R. Di Stefano and S. Mao, APJ, \textbf{93}, 457, (1996).
\end{thebibliography}
\end{document}